\documentclass[pdflatex,sn-mathphys-num]{sn-jnl}

\usepackage{graphicx}%
\usepackage{multirow}%
\usepackage{amsmath,amssymb,amsfonts}%
\usepackage{amsthm}%
\usepackage{mathrsfs}%
\usepackage[title]{appendix}%
\usepackage{xcolor}%
\usepackage{textcomp}%
\usepackage{manyfoot}%
\usepackage{booktabs}%
\usepackage{algorithm}%
\usepackage{algorithmicx}%
\usepackage{algpseudocode}%
\usepackage{listings}%
\usepackage[table]{xcolor}

\usepackage{bm}
\usepackage{color, soul}

\usepackage{amsmath,amssymb,amsfonts}
\usepackage{bm}
\usepackage{color, soul}
\usepackage{graphicx}
\usepackage{float}
\usepackage{wrapfig}
\usepackage{verbatim}
\usepackage[english]{babel}
\usepackage[dvipsnames]{xcolor}
\usepackage{subfigure}
\usepackage{amstext, mathrsfs, textcomp}
\usepackage{multirow}
\usepackage{siunitx}
\usepackage{enumerate}
\usepackage{enumitem}
\usepackage{epstopdf}
\usepackage{tabularx}
\usepackage{lineno}
\makeatletter

\usepackage{physics}
\usepackage{upgreek}

\usepackage{gensymb}

\DeclareSIUnit\sq{\ensuremath\Box}
\DeclareSIUnit\Kelvin{K}
\sethlcolor{yellow}

\graphicspath{{Figures/}}

\theoremstyle{thmstyleone}%
\theoremstyle{thmstyletwo}%

\theoremstyle{thmstylethree}%

\begin{document}

\title[Optical Control of Chirality by Ultrafast Symmetry Breaking in Membrane Metasurfaces]{Optical Control of Chirality by Ultrafast Symmetry Breaking in Membrane Metasurfaces}



\author[1]{\fnm{Nikita} \sur{Glebov}}

\author[2]{\fnm{Alena} \sur{Mamonova}}

\author[3,4]{\fnm{Olesia} \sur{Pashina}}

\author[1]{\fnm{Felix Ulrich} \sur{Brikh}}

\author[3,5]{\fnm{Aleksei} \sur{Ezerskii}}

\author[3,5]{\fnm{Sergey V.} \sur{Makarov}}

\author[3]{\fnm{Mihail} \sur{Petrov}}

\author[2,6]{\fnm{Maxim} \sur{Gorkunov}}

\author*[1]{\fnm{Ivan} \sur{Sinev}}\email{ivan.sinev@epfl.ch}

\author*[1]{\fnm{Hatice} \sur{Altug}}\email{hatice.altug@epfl.ch}

\affil[1]{\orgdiv{Institute of Bioengineering}, \orgname{\'Ecole Polytechnique F\'ed\'erale de Lausanne (EPFL)}, \orgaddress{\city{Lausanne}, \postcode{1015}, \country{Switzerland}}}

\affil[2]{\orgdiv{Shubnikov Institute of Crystallography}, \orgname{NRC “Kurchatov Institute”} \orgaddress{\city{Moscow}, \postcode{119333}, \country{Russia}}}

\affil[3]{\orgdiv{School of Physics and Engineering}, \orgname{ITMO University}, \orgaddress{\city{St. Petersburg}, \postcode{197101}, \country{Russia}}}

\affil[4]{\orgname{University of Brescia}, \orgaddress{\city{Brescia}, \postcode{25121}, \country{Italy}}}

\affil[5]{\orgdiv{Qingdao Innovation and Development Center}, \orgname{Harbin Engineering University}, \orgaddress{\city{Quingdao}, \postcode{266000}, \country{China}}}

\affil[6]{\orgdiv{Theoretical Physics and Quantum Technologies Department}, \orgname{National University of Science and Technology ‘MISIS’}, \orgaddress{\city{Moscow}, \postcode{119049}, \country{Russia}}}

\abstract{Chirality underpins a wide range of light–matter interactions, yet methods for its dynamic control in photonic systems remain limited. Here, we demonstrate ultrafast all-optical control of chirality in silicon metasurfaces through transient symmetry breaking. Our approach exploits photonic eigenstates of opposite spatial parity engineered to be highly susceptible to symmetry perturbations. Optical excitation generates free carriers that establish a transient refractive-index gradient across the membrane thickness, breaking out-of-plane mirror symmetry and facilitating hybridization of the parity-opposite modes into chiral photonic states. This enables the reversible creation and modulation of chirality on a 10--100~ps timescale, manifested by pronounced changes in the metasurface circular dichroism. By dynamically reconfiguring a fundamental symmetry property of the photonic structure rather than merely its optical response, our work establishes a route towards ultrafast control of chiral light–matter interactions and opens opportunities for active nanophotonic and information processing technologies.}

\maketitle

\section*{Main}

Chirality is a fundamental geometric property of objects that cannot be superimposed on their mirror image~\cite{Kelvin1894Clarendon}. It plays a central role in biology, chemistry, and medicine, as many biologically active molecules are inherently chiral~\cite{fasman2013circular, KONDEPUDI20183, RAGHAVAN2018153, kobayashi2011circular}. Likewise, chirality is essential in solid-state physics and quantum photonics, where it underpins a wide range of physical phenomena and applications~\cite{doi:10.1021/acsnano.1c01347, lodahl2017chiral}. The ubiquity of chirality makes its probing crucial from both fundamental and practical perspectives. Light is one of the most efficient probes, as photons  also exhibit chirality in their circularly polarized (CP) states. Asymmetric interaction of CP light with chiral media gives rise to circular dichroism (CD) and optical rotation. However, naturally occurring optical chiral responses are typically weak, so their enhancement is critical for generation and control of chiral light and for sensing applications. Various sophisticated optical structures, including colloidal chiral particles~\cite{ni2023chiral} and complex three-dimensional chiral geometries~\cite{kuzyk2014reconfigurable}, have been developed to enhance chiral light–matter interactions, although their fabrication is often technologically demanding and difficult to scale. Chiral metasurfaces provide an alternative versatile and scalable platform for both enhancing weak natural optical chiral signals and engineering strong artificial optical chirality, enabling precise control and confinement of light and supporting a broad range of optical functionalities~\cite{deng_advances_2024, sinev2025chirality}.
In the near field, they can generate superchiral electromagnetic fields~\cite{doi:10.1126/science.1202817} that enhance interactions with chiral molecules~\cite{ doi:10.1021/acsphotonics.8b00270, doi:10.1021/acsphotonics.8b01767}. From the far-field perspective, their properties are even more unprecedented, as relatively simple structures exhibit the so-called maximum optical chirality~\cite{fernandez-corbaton_objects_2016} when their resonant eigenstates are designed to interact exclusively with waves of a certain helicity~\cite{gorkunov_metasurfaces_2020,kuhner_unlocking_2023}. Apart from their straightforward use for compact optical components performing efficient polarization filtering, conversion, and polarimetric detection~\cite{6529103, doi:10.1021/acsphotonics.2c00395} such resonant eigenstates empower luminescent \cite{han_chiral_2024}, thermal~\cite{sun_circularly_2025} and laser~\cite{deng_chiral_2025} sources of chiral light and even give rise to chiral polaritons by maintaining chiral strong light-matter coupling~\cite{heimig_chiral_2026}.

Establishing efficient dynamic control over strong artificial chirality is the next important challenge in the development of chiral photonic systems with adaptive and reconfigurable functionalities. As optical chirality is inherently related to the geometric chiral properties of the corresponding photonic nanostructure, its substantial variation indeed requires nontrivial approaches. Straightforward ones, based on mechanical reshaping of metasurfaces, have been realized in structures with centimeter-large unit cells operating in the GHz range~\cite{yao_kirigamibased_2026}. For the optical range metasurfaces with sub-micrometer periodicity, recent works employed complex multilayered setups, taking the advantage of chiral Moire patterns formed by differently aligned gratings~\cite{cen_moire_2025} or holey membranes~\cite{Du:26}. However, as their optical chirality arises due to the near-field coupling between the layers, the performance of these implementations critically depends on the ability to sustain submicrometer-thin gaps over the whole areas of the layers, while they rotate and different parts undergo very large relative displacements. The speed and stability of such tunable systems remain determined by the finesse of their micromechanical components and ultimately limits their potential for practical deployment.  More practical non-mechanical approaches rely on combining chiral metasurfaces with tunable optical materials, such as liquid crystals~\cite{ji_active_2021,li_dynamic_2025} and phase-change materials~\cite{sha_chirality_2024}.  Although the absence of moving parts facilitates the tuning stability, the speed is dictated by relatively slow material reconfiguration processes which involve the reorientation of liquid-crystal molecules or the reordering of atomic lattices during phase transitions. 

Here, we report the next pivotal step and experimentally demonstrate that tunable metasurface chirality requires neither mechanical reconfiguration of the nanostructure nor complex combinations of materials. Instead, we control the resonant chirality by addressing the electronic subsystem of the constituent material of the metasurface, enabling an ultrafast and highly reproducible response. By leveraging the high-quality-factor (high-Q) resonances associated with photonic bound states in the continuum (BICs)~\cite{koshelev_engineering_2020}, we realize a system that is extremely sensitive to otherwise very subtle permittivity variations caused by the electron density redistribution~\cite{aigner2025optical}. Our metasurface is based on a free-standing crystalline silicon membrane, which is intrinsically mirror-symmetric with respect to the out-of-plane ($z$) direction, but lacks the in-plane mirror symmetry by the nanostructure design. In our experiment, a femtosecond pulse of visible light impinging the metasurface from one side generates excess charge carriers in silicon close to the corresponding interface. This allows us to establish a transient carrier density profile that induces a strong refractive index gradient in the mid-infrared~\cite{brikh2025mid} and break the remaining optical symmetry plane without any geometric shape transformations. Such optically induced symmetry breaking enables coupling between the engineered modes of opposite spatial parity, leading to their hybridization and the emergence of strong circular dichroism~\cite{GorkunovMaxim_AOM} that decays on a sub-ns timescale governed by the diffusion of the generated charge carriers.  Altogether our study establishes a clear practical route toward ultrafast, low-power, and actively reconfigurable chiral photonic devices.

\begin{figure}[ht]
\centering
\includegraphics[width=\linewidth]{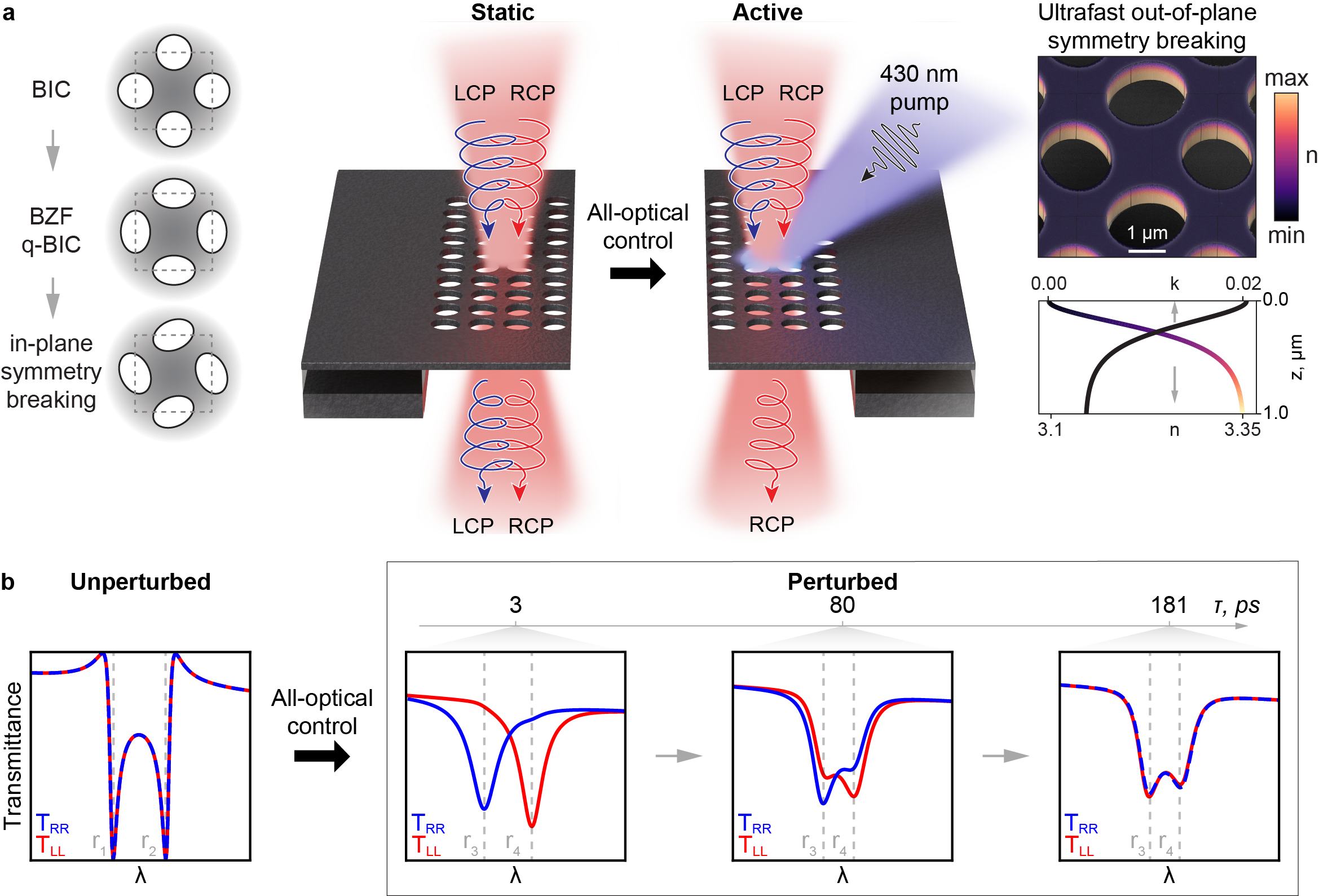}
\caption{\textbf{Concept of optical chirality control by ultrafast symmetry breaking in silicon membrane metasurfaces.} \textbf{a}, The $z$-symmetric silicon membrane metasurface is designed to host quasi-degenerate BZF quasi-BICs of opposite spatial parity 
(see the schematics on the left) exhibiting no optical chirality when probed with circularly polarized mid-infrared light. Upon excitation with a high energy femtosecond pump pulse, optically generated charge carriers induce a pronounced refractive index gradient along the $z$-direction (illustrated with a false color SEM image and a refractive index plot in the right panel), thereby breaking the symmetry and unlocking optical chirality in the mid-infrared range.
\textbf{b}, Temporal evolution of the circularly polarized light transmittance spectra from the achiral unperturbed state (left) through light-induced short-lived strongly chiral (second from the left) to weakly chiral (second from the right) states and then to a nonequilibrium achiral state (right).}
\label{fig1}
\end{figure}

\section*{Hybridized optical states}
For designing metasurfaces with controlled intrinsic optical chirality, we exploit the symmetry properties of photonic eigenstates determined by the metasurface shape symmetry. To avoid polarization conversion which would complicate observations of circular dichroism, we consider metasurfaces of C$_4$ rotational symmetry that forbids polarization conversion of normally incident light. The eigenstates of a metasurface with such symmetry that are coupled to free-space waves are at least double degenerate~\cite{sakoda_symmetry_1995, hopkins_circular_2016, kondratov_extreme_2016, gorkunov_metasurfaces_2020}. For engineering states with high adjustable Q-factors, we employ the concept of Brillouin zone folding (BZF) driven quasi-BICs which are transformed from non-radiating BICs by small shape perturbations increasing the first Brillouin zone of reciprocal space~\cite{overvig_dimerized_2018,wang_brillouin_2023}. In particular, as is shown in left panel of~\autoref{fig1}a, we start from a square array of cylindrical air holes and stretch them in the checkerboard order using the small ellipticity as a parameter explicitly controlling quasi-BIC Q-factors. Rotating the holes breaks all in-plane mirror symmetry planes.

As long as the metasurface out-of-plane mirror symmetry remains intact, its eigenstates can be classified by the spatial parity with respect to the out-of-plane ($z$) direction. We choose two degenerate pairs of quasi-BICs of different parity that remain decoupled from each other due to symmetry constraints. Breaking the vertical symmetry enables their coupling leading to hybridization into chiral quasi-BICs which can manifest themselves in near-maximal chiral optical response even if the symmetry-breaking perturbation is weak, provided that the interacting states are spectrally close~\cite{GorkunovMaxim_AOM}. Below we demonstrate that such strong chiral optical response can be achieved and modulated on ultrafast timescales through the optically induced vertical refractive-index gradients.


To refine the sensitivity to subtle perturbations of dielectric permittivity caused by carrier density inhomogeneity, we use simulations in COMSOL Multiphysics. The initial geometric parameters of the $z$-symmetric metasurface, such as the length, width, and rotation angle of the elliptic holes, as well as the square unit cell size, are optimized to satisfy two conditions. The first one is the spectral position of the resonances. Here, we specifically target the mid-IR range, where strong variations of refractive index are available in silicon for very moderate optical pump fluences~\cite{brikh2025mid}. The second condition is the quasi-degeneracy of the pair of resonances that will be coupled upon breaking the $z$-symmetry. To satisfy it, we adopt a realistic permittivity gradient devised from an extended two-temperature model~\cite{sivan2020ultrafast} and further tune the design to maximize the optical chirality in the perturbed state. The eigenstate circular dichroism is taken as an objective function characterizing its contribution to the observable circular dichroism~\cite{toftul_chiral_2024}, determined by the overlap volume integrals of the displacement current with circularly polarized plane waves. It should be noted that after the final optimization of the perturbed membrane, the eigenfrequencies of the initial fully degenerate resonances of achiral membrane appear to be slightly diverging. This is a prerequisite for achieving maximum chirality in the perturbed state.

The proposed approach to design metasurfaces susceptible to subtle symmetry breaking is illustrated in \autoref{fig1}. The optimized unperturbed ($z$-symmetric) structure exhibits two narrow dips of the transmission down to the zero level, which are underpinned by the photonic eigenstates of different spatial parity that are insensitive to the handedness of their excitation (\autoref{fig1}b left). Upon introducing an optically generated refractive-index gradient (\autoref{fig1}a, right), these modes hybridize, and each of them now is predominantly excited by waves of a particular circular polarization (\autoref{fig1}b, middle). On a sub-ns time scale, the generated charge carriers diffuse, returning the membrane back to the achiral state (\autoref{fig1}b, right).

\section*{Intrinsic and extrinsic chirality in membrane metasurfaces}
\begin{figure}[ht]
\centering
\includegraphics[width=\linewidth]{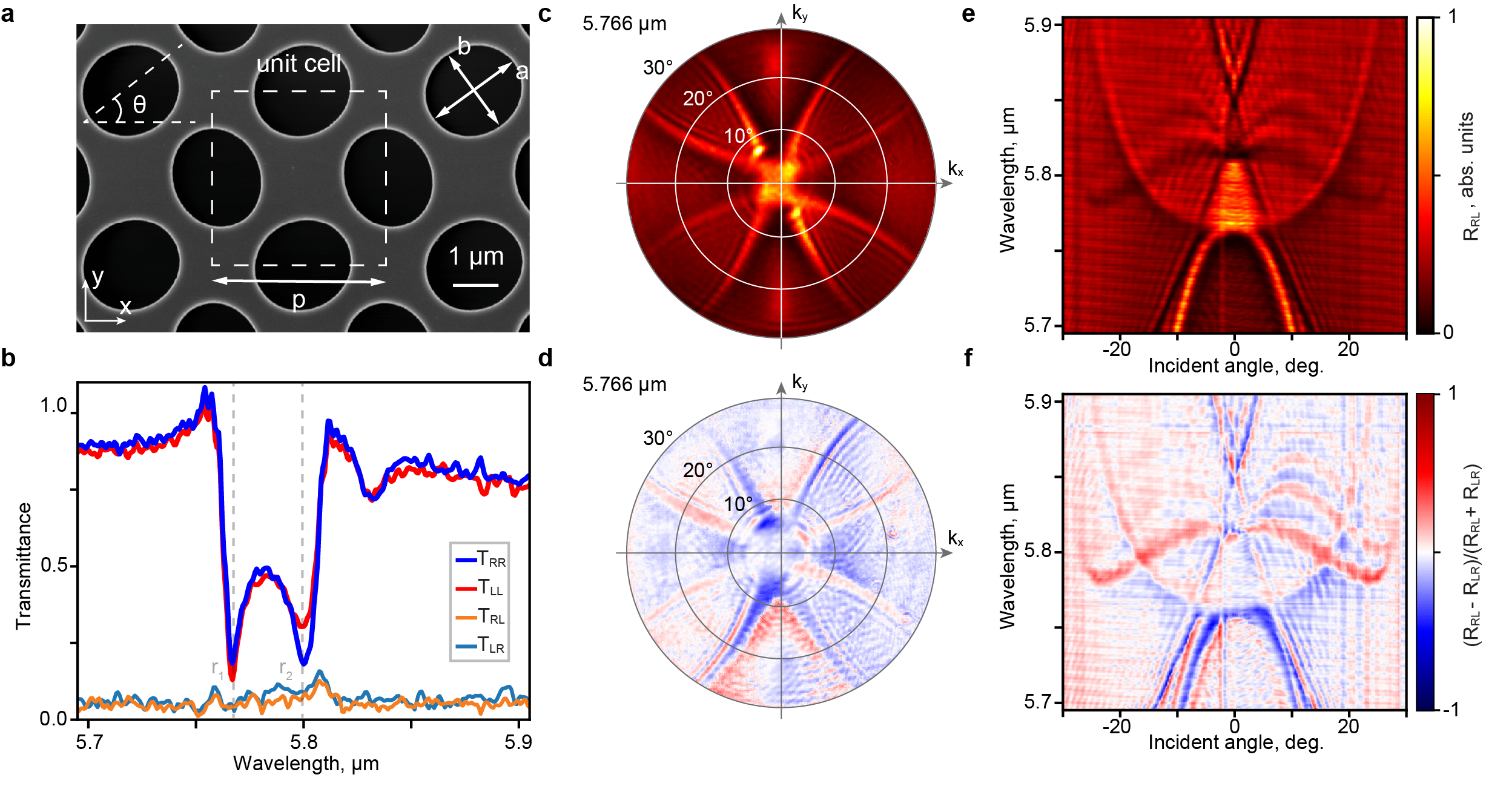}
\caption{\textbf{Intrinsic and extrinsic chirality in membrane metasurfaces.} \textbf{a}, SEM image of the C$_4$-symmetric silicon membrane with unit cell indicated with a dashed square. The design parameters are the primary axes of the elliptical holes (a,b), period p and the ellipse tilt angle $\theta$. \textbf{b}, Circularly polarized transmission spectra of the membrane measured with FTIR at normal incidence.  \textbf{c,d} Back focal plane image of the cross-polarized membrane reflection R$_{RL}$ (c) and conversion CD (d) measured for monochromatic excitation at the resonant wavelength of the higher energy mode. \textbf{e,f}, Angular resolved reflectivity map of R$_{RL}$ and conversion CD, respectively, measured  along the y plane of the metasurface. }
\label{fig2}
\end{figure}

Based on the hybridized-state mechanism described above, suspended mid-infrared silicon membranes provide a unique platform for controlling optical chirality owing to their out-of-plane symmetry and efficient optically induced charge-carrier generation, which enables the formation of substantial refractive-index gradients across the membrane thickness. Furthermore, efficient carrier generation in silicon by visible wavelength pump allows optical excitation to be spatially confined at a scale nearly an order of magnitude smaller than the operating mid-infrared wavelength, opening the way for precise manipulation of the modal structure.  We employ standard CMOS-grade silicon-on-insulator substrates with 1~$\mu$m thick device layer, which possess finite optical losses in the mid-infrared, that can be tuned with doping, and crystalline structure that facilitates high precision processing. This allows for precise control of both radiative and non-radiative optical losses, which is critical for achieving extremely high-Q resonances with strong contrast~\cite{brikh2025mid}. A scanning electron microscopy (SEM) image of the membrane metasurface fabricated with electron beam lithography based on the optimized design is shown in \autoref{fig2}a. The structure has a period of $p = 3.84~\mu\text{m}$, with elliptical holes defined by major and minor axes $a = 2.19~\mu\text{m}$ and $b = 1.93~\mu\text{m}$, respectively, and a tilt angle of $\theta = 24^\circ$.

We start with characterizing linear transmittance spectra of silicon membrane metasurfaces. Optical measurements of the membranes were performed using a customized Fourier-transform infrared (FTIR) microscope accessory equipped with additional polarization optics and refractive ZnSe lenses that enable excitation and analysis in both right- and left-circular polarizations. Additionally, we used back focal plane filtering to restrict the angular spread of the collected light down to an effective numerical aperture (NA) of 0.02, which allowed us to selectively characterize intrinsic chirality of the membrane. 

Transmission spectra measured in co- and cross-circular polarization configurations are presented in \autoref{fig2}b. As expected, the co-polarized transmission components, $T_{RR}$ and $T_{LL}$, exhibit a pair of resonances with large amplitude contrast approaching 80\% and high quality factors of up to $\sim$900, while the cross-polarized components are very small due to the C$_4$ membrane symmetry. Notably, we observe a finite intrinsic CD in the co-polarized components. This highlights the extreme sensitivity of the designed structure to the $z$-symmetry breaking, which in this case arises from minor metasurface sidewall inclination from the etching process. Importantly, as we show later, this residual geometric chirality can be nullified optically by inducing charge carrier gradient.

To characterize the extrinsic chirality of the membrane metasurfaces, we developed a custom mid-infrared polarizaiton-resolved back focal plane (BFP) imaging setup. In this system, the spectrally tunable narrowband quantum cascade laser beam was focused onto the sample surface using a black diamond lens (NA = 0.56), enabling access to incident angles up to $\sim 34^\circ$. The BFP of this optical system was imaged onto a mid-infrared microbolometer camera using a $4f$ optical configuration, allowing direct measurement of angle-resolved reflectivity. By recording these images while tuning the laser wavelength, we obtained polarization- and angle-resolved hyperspectral reflectivity data, which reveal the topology of the isofrequency contours of the metasurface modes. \autoref{fig2}c,d shows the back focal plane images of reflection $R_{RL}$ and conversion circular dichroism~\cite{fedotov2006asymmetric} $\left(R_{RL}-R_{LR}\right)/\left(R_{RL}+R_{LR}\right)$ at a wavelength of $5.765 \ \mu m$ corresponding to the higher energy resonance from the pair (r$_1$ in \autoref{fig2}b). The reflection BFP image generally reproduce the C$_4$ symmetry, however, the conversion CD map shows sign reversal that reduces the symmetry of the image to C$_2$. Collecting the full hyperspectral BFP data allows to further reconstruct the angular dispersion of the metasurface modes (\autoref{fig2}e,f). These maps, plotted from R$_{RL}$ and conversion CD data for the y plane of incidence, reveal two dispersive modes whose chiral response increases with oblique incidence, consistent with the emergence of extrinsic chirality. 

\section*{Chirality by optically induced symmetry breaking}

\begin{figure}[ht]
\centering
\includegraphics[width=\linewidth]{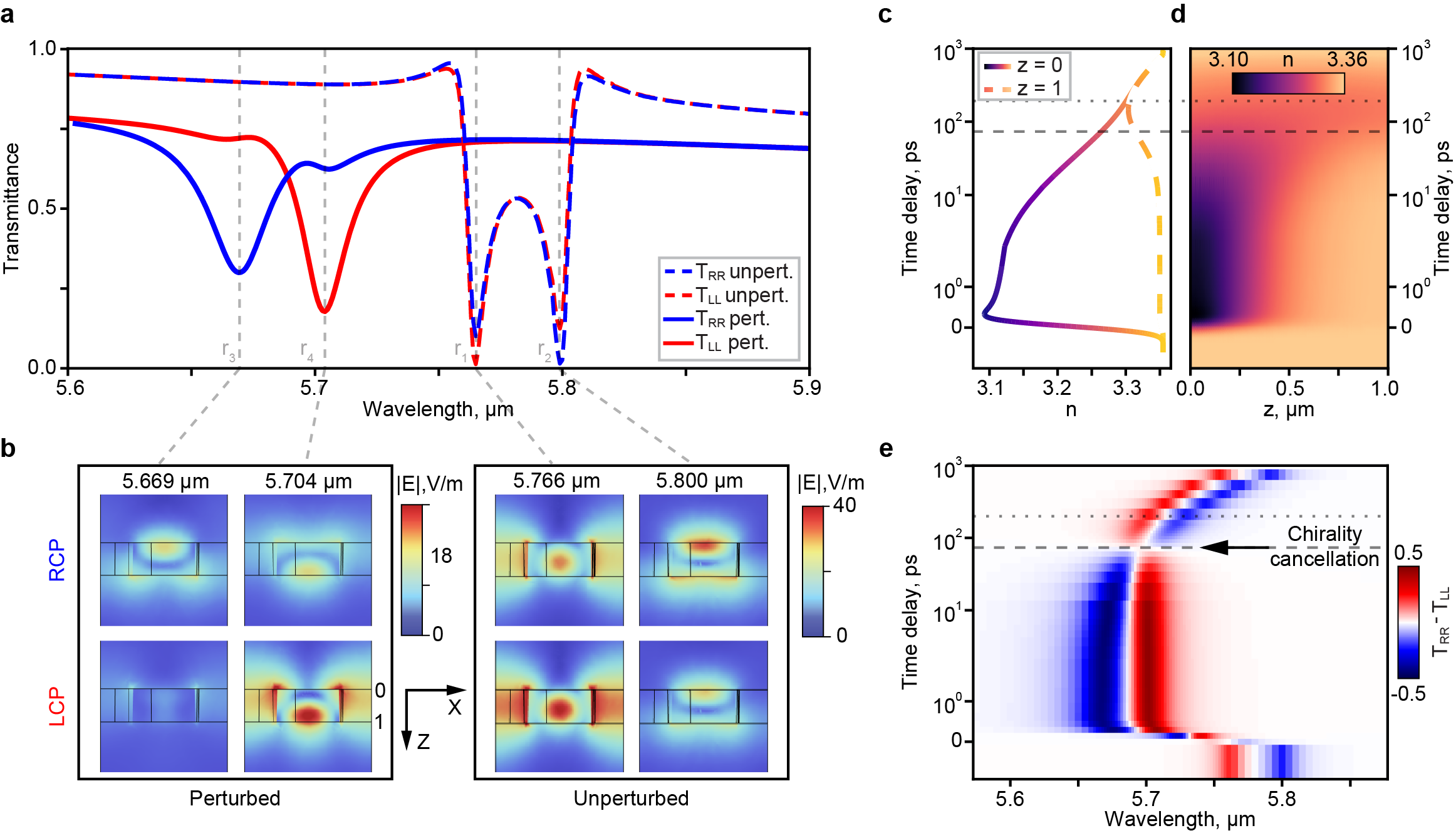}
\caption{\textbf{Inducing chirality by symmetry breaking in suspended membrane metasurfaces.} \textbf{a}, Simulated circularly polarized transmission spectra of the unperturbed (unpert., dashed curves) and the optically perturbed (pert., solid curves) metasurfaces with account for wall slanting. 
\textbf{b}, Field distributions in the $xz$ plane through the center of the metasurface unit cell under circularly polarized excitation of opposite helicities. The maps correspond to the resonances of the unperturbed and perturbed metasurfaces, marked by vertical dashed lines in panel a. \textbf{c}, Temporal evolution of the refractive index of silicon on top (solid curve) and bottom (dashed curve) surfaces of the membrane modified by a short 430~nm laser pulse incident from the top at zero time delay. Horizontal dotted and dashed lines indicate the time points when the optically induced refractive index asymmetry vanishes and compensates geometrical asymmetry, respectively. \textbf{d}, Simulated evolution of the distribution of the refractive index across the membrane thickness. \textbf{e}, Map of the time-resolved transmission contrast $T_{RR}-T_{LL}$ of  the optically pumped metasurface. }

\label{fig3}
\end{figure}

After experimentally verifying the design, we investigate in more detail the mechanism of chirality manipulation \textit{via} optically induced symmetry breaking. Here, we assume a \SI{430}{\nano\meter} pump wavelength, which resides in the spectral range of strong absorption of silicon. This enables efficient generation of charge carriers and steep gradients of their spatial concentration profiles driven by low propagation length of radiation in the material. The pump, incident on the membrane from one of the sides, thus creates a strongly depth-dependent population of nonequilibrium carriers that induce a corresponding refractive index profile.


We calculated the full temporal evolution of this carrier distribution using the extended two-temperature model~\cite{sivan2020ultrafast}. The model accounts for optical carrier generation, relaxation of the initially nonthermal electron--hole population, carrier diffusion and recombination, and energy exchange between the electronic and lattice subsystems. It provides the local time-dependent carrier concentration across the membrane thickness, \(N_e(z,t)\), which is then used to determine the transient optical constants of silicon.

The corresponding dielectric permittivity function depends on the local carrier density. At the probe frequency, it is described by including the free-carrier Drude response together with the corrections associated with band filling and band-gap renormalization:
\begin{equation}
\varepsilon(\omega,z,t)
=
\varepsilon_{\mathrm{Si}}(\omega,N_{e,0})
+
\Delta\varepsilon_{\mathrm{D}}(\omega,N_e)
+
\Delta\varepsilon_{\mathrm{BF}}(\omega,N_e)
+
\Delta\varepsilon_{\mathrm{BGR}}(\omega,N_e).
\end{equation}
Here, \(\varepsilon_{\mathrm{Si}}(\omega,N_{e,0})\) is the dielectric permittivity function of the initially doped silicon membrane, \(\Delta\varepsilon_{\mathrm{D}}\) is the Drude contribution of free carriers, while \(\Delta\varepsilon_{\mathrm{BF}}\) and \(\Delta\varepsilon_{\mathrm{BGR}}\) account for band filling and band-gap renormalization, respectively~\cite{sokolowski2000generation}.

At the mid-infrared probe wavelength, the photon energy is far below the silicon band gap. Therefore, the refractive-index modulation is governed mainly by the free-carrier Drude response. The dominant carrier-induced correction can be written as
\begin{equation}
\Delta\varepsilon_{\mathrm{D}}(\omega,N_e)
=
-\frac{N_e e^2}{\varepsilon_0 m_{\mathrm{opt}}^{\ast}}
\frac{1}{\omega(\omega+i/\tau_{\mathrm{D}})} ,
\end{equation}
where \(m_{\mathrm{opt}}^{\ast}\) is the optical effective mass of the electron--hole plasma and \(\tau_{\mathrm{D}}\) is the momentum relaxation time. This correction reduces the real part of the dielectric function and increases optical losses. As the strength of the Drude correction is set by the local carrier density, the nonuniform \(N_e(z,t)\) profile produced by the pump leads to a pronounced spatially varying dielectric response at the probe wavelength. The highest carrier density right after the arrival of the pulse is reached near the illuminated surface, where \(n\) experiences the strongest decrease and \(k\) increases. This evolution, calculated using the known optical constants of silicon and a pump intensity of $9\cdot10^{12}$\SI{}{\watt\per\meter\squared}, is shown in \autoref{fig3}c,d. 

We then used the estimated refractive index profiles to calculate the corresponding polarization-resolved transmission spectra. As shown before, the co-polarized transmittances $T_{RR}$ and $T_{LL}$ of the fabricated membrane are slightly different due to geometrical symmetry breaking (\autoref{fig2}b). This is perfectly reproduced in the simulation spectra where we introduced a 1 degree slant angle to the membrane walls (\autoref{fig3}a, dashed curves). Introduction of the calculated refractive-index gradient leads to a blue shift of the resonant modes and the emergence of a strong circular polarized transmission contrast. The value of circular dichroism, that we define here as $\text{CD} = \left({T_{RR} - T_{LL}}\right)/\left({T_{RR} + T_{LL}}\right)$, reaches 0.56 at the resonant modes.

To further illustrate how strong chirality emerges through mode hybridization, we calculate the resonant electric-field distributions within the membrane. The eigenstate simulations of a z-symmetric membrane confirm different spatial parities in the out-of-plane direction required for further hybridization. \autoref{fig3}b shows the field distributions in the $xz$ plane at the center of the realistic membrane for RCP and LCP excitations for both the maximally detuned resonances (r$_3$,r$_4$) and the original resonances (r$_1$, r$_2$).  In the unperturbed membrane, the field distributions are almost identical for RCP and LCP excitation, with minor differences emerging due to a finite sidewall angle (\autoref{fig3}b, right). In contrast, in the perturbed membrane, the modes are efficiently hybridized, leading to strong selectivity of the mode field to the excitation helicity (\autoref{fig3}b, left).

The temporal evolution of the optically induced refractive index gradient is governed by two main factors: the diffusion of the non-equilibrium charge carriers and their decay through recombination. Our model allowed us to reconstruct the resulting dynamics of the refractive index distribution, which is shown in \autoref{fig3}c,d.
The data show a sharp decrease in the refractive index occuring at the top surface upon pump arrival. In contrast, the refractive index at the bottom surface changes more gradually, decreasing as carriers diffuse through the membrane. As a result, the refractive-index gradient across the thickness diminishes and eventually vanishes. After that, the dynamics are defined  exclusively by the charge recombination process, which has a characteristic time of a few ns~\cite{brikh2025mid}.

This process directly contributes to the dynamic modification of the metasurface chirality. \autoref{fig3}e shows the time-resolved spectra of the circularly polarized transmission contrast $T_{RR}-T_{LL}$. Before the arrival of the pump pulse at $t=0$, the membrane exhibits weak chirality due to the sidewall slanting. At zero time delay, the transmission contrast increases drastically; furthermore, it flips its sign, which is indicative of optically induced chirality compensating and overpowering the geometric factor. Over the next few hundreds of ps, diffusion of the generated charge carriers gradually erases the contrast. At 70~ps the optical and geometric factors compensate each other perfectly, rendering the membrane achiral. Optically induced asymmetry becomes negligible ($<1\%$ of geometric factor) at $t\approx$450~ps, after which the perturbed resonances relax to their original spectral position \textit{via} concurrent charge recombination without changing of chirality.

\section*{Ultrafast all-optical chirality control}

\begin{figure}[ht!]
\centering
\includegraphics[width=\linewidth]{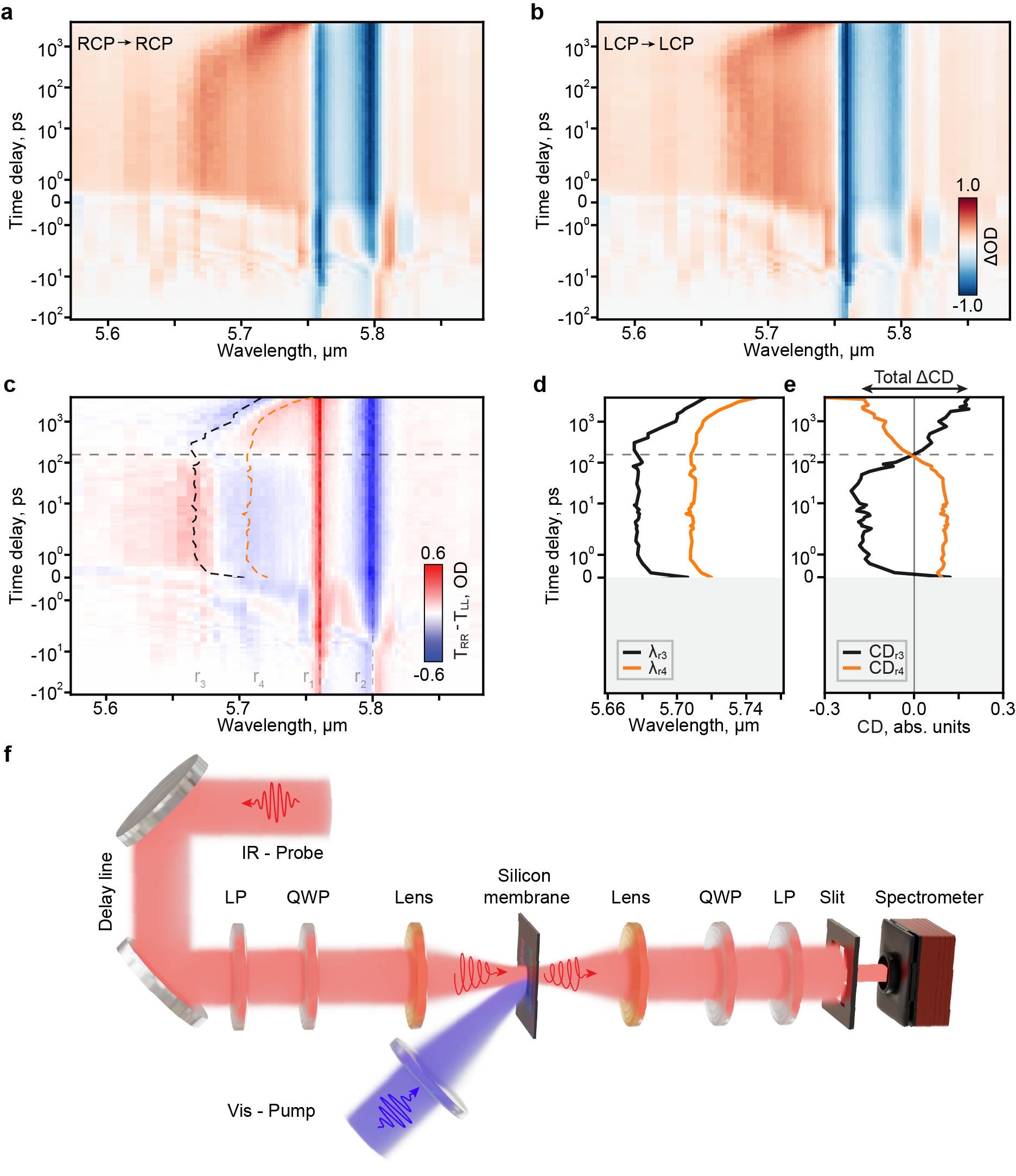}
\caption{\textbf{Ultrafast all-optical chirality control in nonlinear experiments.}  \textbf{a}, Transient absorption spectroscopy map of the membrane extracted from co-polarized RCP transmission. \textbf{b}, Map for co-polarized LCP transmission. \textbf{c}, Differential absorption map calculated from data in panels c and d. Spectral positions of  original resonances ($r_1$,$r_2$) and the maximally detuned resonances ($r_3$,$r_4$) are marked with dashed lines. \textbf{d},  Temporal evolution of the wavelengths of the detuned resonances ($r_3$,$r_4$) fitted from the map in panel d. \textbf{e}, Temporal evolution of circular dichroism  at the resonant wavelengths of $r_3$ and $r_4$. \textbf{f}, Schematic of the mid-infrared circularly polarized pump-probe spectroscopy.}
\label{fig4}
\end{figure}

To experimentally demonstrate ultrafast control of chirality in our membrane metasurface, we employ mid-IR transient absorption spectroscopy. Visible femtosecond light pulses (430~nm) are used to generate a gradient of charge carriers within the membrane, and we track the temporal evolution of the chirality by probing the transmission in co-circular polarized channels with time-delayed spectrally tunable mid-infrared pulses (see \autoref{fig4}f). Using this setup, we measured the time-resolved spectra of differential optical density ($\Delta \mathrm{OD}$) for co-circular polarized transmission at a pump fluence of~\SI{19.2}{\micro\joule\per\centi\meter\squared}(\autoref{fig4}a,b). Here, the optical density is defined as
\begin{equation*}
    \Delta \text{OD}(\lambda, t) = \log_{10}\left( \frac{T^{u}}{T^{p}} \right),
\end{equation*}
where $\tau$ is the pump–probe delay, and $T^{u}$ and $T^{p}$ are the transmission signals of the unperturbed and perturbed membranes, respectively.  

The differential OD spectra (\autoref{fig4}a,b) exhibit two dips at longer wavelengths corresponding to the original resonances of the unperturbed metasurface (r$_1$, r$_2$). These dips are weakly manifested even before pump arrival due to residual heating of the membrane, as the limited heat dissipation in a large freestanding membrane prevents full relaxation between successive pulses. Furthermore, in the experimental maps we observe fringes spanning from \SI{5.6}{\micro\meter} to \SI{5.8}{\micro\meter} at negative time delays, which are indicative of frequency conversion of probe beam coupled to the metasurface\cite{KarlNanoLett}. This effect is attributed to the interference  between a long-lived resonant coupled mode and its dynamically modified counterpart induced by the subsequent optical pump.

Upon pump arrival, a blue-shifted peak appears in each of the co-polarized OD maps. For both polarizations, it subsequently relaxes back to merge with the dips that correspond to its original spectral position. However, the temporal reshaping of this peak has pronounced differences between RCP and LCP data, which is the signature of dynamically changing chirality. To quantify it, we evaluate the difference between the co-circular polarization signals (\autoref{fig4}c). In these data, finite CD is observed at the original resonance positions $r_1$ and $r_2$, consistent with the FTIR measurements and attributed to minor etching-related slanting of the membrane walls. Following the pump arrival at t=0, a pronounced transient CD emerges at the blue-shifted resonances $r_3$ and $r_4$. It is important to note that identical values of transmission contrast at the original and detuned resonances correspond to flipping of the sign of chirality: by the same logic, in the differential OD maps (\autoref{fig4}a,b) the resonant transmission dips manifest as minima at the original positions and as maxima after detuning. Therefore, the  contrast in \autoref{fig4}c indicates that the optical symmetry breaking introduced by the optical pump is enough not only to compensate, but reverse the original chirality of the metasurface in full agreement in the simulation (\autoref{fig3}e). As the induced refractive index gradient dissipates according to the dynamics shown in \autoref{fig3}d, the membrane becomes achiral at t$\simeq$150~ps. This corresponds to a perfect compensation of residual geometric chirality and optically induced one.  

While the differential OD data provides qualitative understanding of the chirality dynamics, it needs to be further processed to extract quantitative values such as CD. To do that, we extracted the resonance wavelengths from the difference between optical density maps of the co-circular polarizations (in \autoref{fig4}c) by fitting the data with a sum of four Lorentzian functions, corresponding to four resonances: two blue-shifted modes ($r_3$ and $r_4$) and two original modes ($r_1$ and $r_2$).  The resonant wavelengths of the perturbed structure $\lambda_{r3}$ and $\lambda_{r4}$ initially exhibit a blue shift, followed by a plateau during carrier diffusion and concentration leveling, and subsequently red-shift toward their original positions as the charge carriers recombine (\autoref{fig4}d).
 
\autoref{fig4}e shows time-dependent CD, calculated as a function of the pump--probe delay. Following the optical excitation, a transient CD emerges, reversing the initial residual geometric chirality, and then decaying on a timescale of a few hundred of ps in excellent agreement with theoretical calculations (see \autoref{fig3}e). As the carrier distribution becomes homogenized across the membrane thickness, the CD gradually returns toward its equilibrium magnitude with opposite sign. The maximum optically induced CD reaches $\sim$0.2, whereas the overall CD modulation approaches $\sim$0.4, in good agreement with theoretical predictions.
   
\section*{Conclusion}
To conclude, we experimentally demonstrate ultrafast optical control of chirality in metasurfaces enabled by temporal symmetry breaking. The key to this approach is the implementation of highly sensitive photonic eigenstates of opposite spatial parity in the mid-infrared, where optically induced charge carriers generate strong refractive-index gradients at moderate pump fluences. This enables transient symmetry breaking driven by a visible pump, leading to dynamic, large amplitude tuning of CD and even compensation of residual geometric asymmetry. We experimentally achieve modulation depths of $\mathrm{CD}\sim0.4$, including the creation and annihilation of chiral response on~$10-100$~ps timescales, in agreement with modeling based on an extended two-temperature framework describing carrier dynamics. Furthermore, both the temporal response and spectral position of the induced circular dichroism can be controlled by the pump fluence. 

The ability to dynamically control the fundamental symmetry of metasurfaces establishes a powerful paradigm in active nanophotonics, enabling reconfigurable polarization control and tunable light–matter interactions. Moreover, all-optical control of chirality paves the way for advanced functionalities in high-speed, low-power optical communication, information encoding, applications in nonlinear photonics, as well as chemical and biological sensing with enhanced symmetry sensitivity.

\newpage


\bibliography{sn-bibliography}


\begin{thebibliography}{43}
\ifx \bisbn   \undefined \def \bisbn  #1{ISBN #1}\fi
\ifx \binits  \undefined \def \binits#1{#1}\fi
\ifx \bauthor  \undefined \def \bauthor#1{#1}\fi
\ifx \batitle  \undefined \def \batitle#1{#1}\fi
\ifx \bjtitle  \undefined \def \bjtitle#1{#1}\fi
\ifx \bvolume  \undefined \def \bvolume#1{\textbf{#1}}\fi
\ifx \byear  \undefined \def \byear#1{#1}\fi
\ifx \bissue  \undefined \def \bissue#1{#1}\fi
\ifx \bfpage  \undefined \def \bfpage#1{#1}\fi
\ifx \blpage  \undefined \def \blpage #1{#1}\fi
\ifx \burl  \undefined \def \burl#1{\textsf{#1}}\fi
\ifx \doiurl  \undefined \def \doiurl#1{\url{https://doi.org/#1}}\fi
\ifx \betal  \undefined \def \betal{\textit{et al.}}\fi
\ifx \binstitute  \undefined \def \binstitute#1{#1}\fi
\ifx \binstitutionaled  \undefined \def \binstitutionaled#1{#1}\fi
\ifx \bctitle  \undefined \def \bctitle#1{#1}\fi
\ifx \beditor  \undefined \def \beditor#1{#1}\fi
\ifx \bpublisher  \undefined \def \bpublisher#1{#1}\fi
\ifx \bbtitle  \undefined \def \bbtitle#1{#1}\fi
\ifx \bedition  \undefined \def \bedition#1{#1}\fi
\ifx \bseriesno  \undefined \def \bseriesno#1{#1}\fi
\ifx \blocation  \undefined \def \blocation#1{#1}\fi
\ifx \bsertitle  \undefined \def \bsertitle#1{#1}\fi
\ifx \bsnm \undefined \def \bsnm#1{#1}\fi
\ifx \bsuffix \undefined \def \bsuffix#1{#1}\fi
\ifx \bparticle \undefined \def \bparticle#1{#1}\fi
\ifx \barticle \undefined \def \barticle#1{#1}\fi
\bibcommenthead
\ifx \bconfdate \undefined \def \bconfdate #1{#1}\fi
\ifx \botherref \undefined \def \botherref #1{#1}\fi
\ifx \url \undefined \def \url#1{\textsf{#1}}\fi
\ifx \bchapter \undefined \def \bchapter#1{#1}\fi
\ifx \bbook \undefined \def \bbook#1{#1}\fi
\ifx \bcomment \undefined \def \bcomment#1{#1}\fi
\ifx \oauthor \undefined \def \oauthor#1{#1}\fi
\ifx \citeauthoryear \undefined \def \citeauthoryear#1{#1}\fi
\ifx \endbibitem  \undefined \def \endbibitem {}\fi
\ifx \bconflocation  \undefined \def \bconflocation#1{#1}\fi
\ifx \arxivurl  \undefined \def \arxivurl#1{\textsf{#1}}\fi
\csname PreBibitemsHook\endcsname

\bibitem[\protect\citeauthoryear{Kelvin}{}]{Kelvin1894Clarendon}
\begin{botherref}
\oauthor{\bsnm{Kelvin}, \binits{W.T.}}:
The Molecular Tactics of a Crystal.
Robert Boyle Lecture, 1893.
Clarendon Press
\end{botherref}
\endbibitem

\bibitem[\protect\citeauthoryear{Fasman}{}]{fasman2013circular}
\begin{botherref}
\oauthor{\bsnm{Fasman}, \binits{G.D.}}:
Circular Dichroism and the Conformational Analysis of Biomolecules.
Springer
\end{botherref}
\endbibitem

\bibitem[\protect\citeauthoryear{Kondepudi}{2018}]{KONDEPUDI20183}
\begin{bchapter}
\bauthor{\bsnm{Kondepudi}, \binits{D.}}:
\bctitle{Chapter 1 - chiral asymmetry in nature}.
In: \beditor{\bsnm{Polavarapu}, \binits{P.L.}} (ed.)
\bbtitle{Chiral Analysis},
pp. \bfpage{3}--\blpage{28}.
\bpublisher{Elsevier},
\blocation{Amsterdam}
(\byear{2018}).
\doiurl{10.1016/B978-0-444-64027-7.00001-X}
\end{bchapter}
\endbibitem

\bibitem[\protect\citeauthoryear{Raghavan and Polavarapu}{2018}]{RAGHAVAN2018153}
\begin{bchapter}
\bauthor{\bsnm{Raghavan}, \binits{V.}},
\bauthor{\bsnm{Polavarapu}, \binits{P.L.}}:
\bctitle{Chapter 4 - chiroptical spectroscopic studies on soft aggregates and their interactions}.
In: \bbtitle{Chiral Analysis},
pp. \bfpage{153}--\blpage{200}.
\bpublisher{Elsevier},
\blocation{Amsterdam}
(\byear{2018}).
\doiurl{10.1016/B978-0-444-64027-7.00004-5}
\end{bchapter}
\endbibitem

\bibitem[\protect\citeauthoryear{Kobayashi et~al.}{2011}]{kobayashi2011circular}
\begin{bbook}
\bauthor{\bsnm{Kobayashi}, \binits{N.}},
\bauthor{\bsnm{Muranaka}, \binits{A.}},
\bauthor{\bsnm{Mack}, \binits{J.}}:
\bbtitle{Circular Dichroism and Magnetic Circular Dichroism Spectroscopy for Organic Chemists}.
\bpublisher{Royal Society of Chemistry},
\blocation{Cambridge}
(\byear{2011}).
\doiurl{10.1039/9781849732932}
\end{bbook}
\endbibitem

\bibitem[\protect\citeauthoryear{Aiello et~al.}{2022}]{doi:10.1021/acsnano.1c01347}
\begin{barticle}
\bauthor{\bsnm{Aiello}, \binits{C.D.}},
\bauthor{\bsnm{Abendroth}, \binits{J.M.}},
\bauthor{\bsnm{Abbas}, \binits{M.}},
\bauthor{\bsnm{Afanasev}, \binits{A.}},
\bauthor{\bsnm{Agarwal}, \binits{S.}},
\bauthor{\bsnm{Banerjee}, \binits{A.S.}},
\bauthor{\bsnm{Beratan}, \binits{D.N.}},
\bauthor{\bsnm{Belling}, \binits{J.N.}},
\bauthor{\bsnm{Berche}, \binits{B.}},
\bauthor{\bsnm{Botana}, \binits{A.}},
\bauthor{\bsnm{Caram}, \binits{J.R.}},
\bauthor{\bsnm{Celardo}, \binits{G.L.}},
\bauthor{\bsnm{Cuniberti}, \binits{G.}},
\bauthor{\bsnm{Garcia-Etxarri}, \binits{A.}},
\bauthor{\bsnm{Dianat}, \binits{A.}},
\bauthor{\bsnm{Diez-Perez}, \binits{I.}},
\bauthor{\bsnm{Guo}, \binits{Y.}},
\bauthor{\bsnm{Gutierrez}, \binits{R.}},
\bauthor{\bsnm{Herrmann}, \binits{C.}},
\bauthor{\bsnm{Hihath}, \binits{J.}},
\bauthor{\bsnm{Kale}, \binits{S.}},
\bauthor{\bsnm{Kurian}, \binits{P.}},
\bauthor{\bsnm{Lai}, \binits{Y.-C.}},
\bauthor{\bsnm{Liu}, \binits{T.}},
\bauthor{\bsnm{Lopez}, \binits{A.}},
\bauthor{\bsnm{Medina}, \binits{E.}},
\bauthor{\bsnm{Mujica}, \binits{V.}},
\bauthor{\bsnm{Naaman}, \binits{R.}},
\bauthor{\bsnm{Noormandipour}, \binits{M.}},
\bauthor{\bsnm{Palma}, \binits{J.L.}},
\bauthor{\bsnm{Paltiel}, \binits{Y.}},
\bauthor{\bsnm{Petuskey}, \binits{W.}},
\bauthor{\bsnm{Ribeiro-Silva}, \binits{J.C.}},
\bauthor{\bsnm{Saenz}, \binits{J.J.}},
\bauthor{\bsnm{Santos}, \binits{E.J.G.}},
\bauthor{\bsnm{Solyanik-Gorgone}, \binits{M.}},
\bauthor{\bsnm{Sorger}, \binits{V.J.}},
\bauthor{\bsnm{Stemer}, \binits{D.M.}},
\bauthor{\bsnm{Ugalde}, \binits{J.M.}},
\bauthor{\bsnm{Valdes-Curiel}, \binits{A.}},
\bauthor{\bsnm{Varela}, \binits{S.}},
\bauthor{\bsnm{Waldeck}, \binits{D.H.}},
\bauthor{\bsnm{Wasielewski}, \binits{M.R.}},
\bauthor{\bsnm{Weiss}, \binits{P.S.}},
\bauthor{\bsnm{Zacharias}, \binits{H.}},
\bauthor{\bsnm{Wang}, \binits{Q.H.}}:
\batitle{A chirality-based quantum leap}.
\bjtitle{ACS Nano}
\bvolume{16}(\bissue{4}),
\bfpage{4989}--\blpage{5035}
(\byear{2022})
\doiurl{10.1021/acsnano.1c01347} .
\bcomment{PMID: 35318848}
\end{barticle}
\endbibitem

\bibitem[\protect\citeauthoryear{Lodahl et~al.}{2017}]{lodahl2017chiral}
\begin{barticle}
\bauthor{\bsnm{Lodahl}, \binits{P.}},
\bauthor{\bsnm{Mahmoodian}, \binits{S.}},
\bauthor{\bsnm{Stobbe}, \binits{S.}},
\bauthor{\bsnm{Rauschenbeutel}, \binits{A.}},
\bauthor{\bsnm{Schneeweiss}, \binits{P.}},
\bauthor{\bsnm{Volz}, \binits{J.}},
\bauthor{\bsnm{Pichler}, \binits{H.}},
\bauthor{\bsnm{Zoller}, \binits{P.}}:
\batitle{Chiral quantum optics}.
\bjtitle{Nature}
\bvolume{541}(\bissue{7638}),
\bfpage{473}--\blpage{480}
(\byear{2017})
\doiurl{10.1038/nature21037}
\end{barticle}
\endbibitem

\bibitem[\protect\citeauthoryear{Ni et~al.}{2023}]{ni2023chiral}
\begin{barticle}
\bauthor{\bsnm{Ni}, \binits{B.}},
\bauthor{\bsnm{Mychinko}, \binits{M.}},
\bauthor{\bsnm{Gómez-Graña}, \binits{S.}},
\bauthor{\bsnm{Morales-Vidal}, \binits{J.}},
\bauthor{\bsnm{Obelleiro-Liz}, \binits{M.}},
\bauthor{\bsnm{Heyvaert}, \binits{W.}},
\bauthor{\bsnm{Vila-Liarte}, \binits{D.}},
\bauthor{\bsnm{Zhuo}, \binits{X.}},
\bauthor{\bsnm{Albrecht}, \binits{W.}},
\bauthor{\bsnm{Zheng}, \binits{G.}},
\bauthor{\bsnm{González-Rubio}, \binits{G.}},
\bauthor{\bsnm{Taboada}, \binits{J.M.}},
\bauthor{\bsnm{Obelleiro}, \binits{F.}},
\bauthor{\bsnm{López}, \binits{N.}},
\bauthor{\bsnm{Pérez-Juste}, \binits{J.}},
\bauthor{\bsnm{Pastoriza-Santos}, \binits{I.}},
\bauthor{\bsnm{Cölfen}, \binits{H.}},
\bauthor{\bsnm{Bals}, \binits{S.}},
\bauthor{\bsnm{Liz-Marzán}, \binits{L.M.}}:
\batitle{Chiral seeded growth of gold nanorods into fourfold twisted nanoparticles with plasmonic optical activity}.
\bjtitle{Advanced Materials}
\bvolume{35}(\bissue{1}),
\bfpage{2208299}
(\byear{2023})
\doiurl{10.1002/adma.202208299}
\end{barticle}
\endbibitem

\bibitem[\protect\citeauthoryear{Kuzyk et~al.}{2014}]{kuzyk2014reconfigurable}
\begin{barticle}
\bauthor{\bsnm{Kuzyk}, \binits{A.}},
\bauthor{\bsnm{Schreiber}, \binits{R.}},
\bauthor{\bsnm{Zhang}, \binits{H.}},
\bauthor{\bsnm{Govorov}, \binits{A.O.}},
\bauthor{\bsnm{Liedl}, \binits{T.}},
\bauthor{\bsnm{Liu}, \binits{N.}}:
\batitle{Reconfigurable 3d plasmonic metamolecules}.
\bjtitle{Nature materials}
\bvolume{13}(\bissue{9}),
\bfpage{862}--\blpage{866}
(\byear{2014})
\doiurl{10.1038/nmat4031}
\end{barticle}
\endbibitem

\bibitem[\protect\citeauthoryear{Deng et~al.}{2024}]{deng_advances_2024}
\begin{barticle}
\bauthor{\bsnm{Deng}, \binits{Q.-M.}},
\bauthor{\bsnm{Li}, \binits{X.}},
\bauthor{\bsnm{Hu}, \binits{M.-X.}},
\bauthor{\bsnm{Li}, \binits{F.-J.}},
\bauthor{\bsnm{Li}, \binits{X.}},
\bauthor{\bsnm{Deng}, \binits{Z.-L.}}:
\batitle{Advances on broadband and resonant chiral metasurfaces}.
\bjtitle{npj Nanophotonics}
\bvolume{1}(\bissue{1}),
\bfpage{20}
(\byear{2024})
\doiurl{10.1038/s44310-024-00018-5}
\end{barticle}
\endbibitem

\bibitem[\protect\citeauthoryear{Sinev et~al.}{2025}]{sinev2025chirality}
\begin{barticle}
\bauthor{\bsnm{Sinev}, \binits{I.}},
\bauthor{\bsnm{Richter}, \binits{F.U.}},
\bauthor{\bsnm{Toftul}, \binits{I.}},
\bauthor{\bsnm{Glebov}, \binits{N.}},
\bauthor{\bsnm{Koshelev}, \binits{K.}},
\bauthor{\bsnm{Hwang}, \binits{Y.}},
\bauthor{\bsnm{Lancaster}, \binits{D.G.}},
\bauthor{\bsnm{Kivshar}, \binits{Y.}},
\bauthor{\bsnm{Altug}, \binits{H.}}:
\batitle{Chirality encoding in resonant metasurfaces governed by lattice symmetries}.
\bjtitle{nature communications}
\bvolume{16}(\bissue{1}),
\bfpage{6091}
(\byear{2025})
\doiurl{10.1038/s41467-025-61221-2}
\end{barticle}
\endbibitem

\bibitem[\protect\citeauthoryear{Tang and Cohen}{2011}]{doi:10.1126/science.1202817}
\begin{barticle}
\bauthor{\bsnm{Tang}, \binits{Y.}},
\bauthor{\bsnm{Cohen}, \binits{A.E.}}:
\batitle{Enhanced enantioselectivity in excitation of chiral molecules by superchiral light}.
\bjtitle{Science}
\bvolume{332}(\bissue{6027}),
\bfpage{333}--\blpage{336}
(\byear{2011})
\doiurl{10.1126/science.1202817}
{\href{https://arxiv.org/abs/https://www.science.org/doi/pdf/10.1126/science.1202817}{{https://www.science.org/doi/pdf/10.1126/science.1202817}}}
\end{barticle}
\endbibitem

\bibitem[\protect\citeauthoryear{Mohammadi et~al.}{2018}]{doi:10.1021/acsphotonics.8b00270}
\begin{barticle}
\bauthor{\bsnm{Mohammadi}, \binits{E.}},
\bauthor{\bsnm{Tsakmakidis}, \binits{K.L.}},
\bauthor{\bsnm{Askarpour}, \binits{A.N.}},
\bauthor{\bsnm{Dehkhoda}, \binits{P.}},
\bauthor{\bsnm{Tavakoli}, \binits{A.}},
\bauthor{\bsnm{Altug}, \binits{H.}}:
\batitle{Nanophotonic platforms for enhanced chiral sensing}.
\bjtitle{ACS Photonics}
\bvolume{5}(\bissue{7}),
\bfpage{2669}--\blpage{2675}
(\byear{2018})
\doiurl{10.1021/acsphotonics.8b00270}
{\href{https://arxiv.org/abs/https://doi.org/10.1021/acsphotonics.8b00270}{{https://doi.org/10.1021/acsphotonics.8b00270}}}
\end{barticle}
\endbibitem

\bibitem[\protect\citeauthoryear{Mohammadi et~al.}{2019}]{doi:10.1021/acsphotonics.8b01767}
\begin{barticle}
\bauthor{\bsnm{Mohammadi}, \binits{E.}},
\bauthor{\bsnm{Tavakoli}, \binits{A.}},
\bauthor{\bsnm{Dehkhoda}, \binits{P.}},
\bauthor{\bsnm{Jahani}, \binits{Y.}},
\bauthor{\bsnm{Tsakmakidis}, \binits{K.L.}},
\bauthor{\bsnm{Tittl}, \binits{A.}},
\bauthor{\bsnm{Altug}, \binits{H.}}:
\batitle{Accessible superchiral near-fields driven by tailored electric and magnetic resonances in all-dielectric nanostructures}.
\bjtitle{ACS Photonics}
\bvolume{6}(\bissue{8}),
\bfpage{1939}--\blpage{1946}
(\byear{2019})
\doiurl{10.1021/acsphotonics.8b01767}
{\href{https://arxiv.org/abs/https://doi.org/10.1021/acsphotonics.8b01767}{{https://doi.org/10.1021/acsphotonics.8b01767}}}
\end{barticle}
\endbibitem

\bibitem[\protect\citeauthoryear{Fernandez-Corbaton et~al.}{2016}]{fernandez-corbaton_objects_2016}
\begin{barticle}
\bauthor{\bsnm{Fernandez-Corbaton}, \binits{I.}},
\bauthor{\bsnm{Fruhnert}, \binits{M.}},
\bauthor{\bsnm{Rockstuhl}, \binits{C.}}:
\batitle{Objects of {Maximum} {Electromagnetic} {Chirality}}.
\bjtitle{Physical Review X}
\bvolume{6}(\bissue{3}),
\bfpage{031013}
(\byear{2016})
\doiurl{10.1103/PhysRevX.6.031013}
\end{barticle}
\endbibitem

\bibitem[\protect\citeauthoryear{Gorkunov et~al.}{2020}]{gorkunov_metasurfaces_2020}
\begin{barticle}
\bauthor{\bsnm{Gorkunov}, \binits{M.V.}},
\bauthor{\bsnm{Antonov}, \binits{A.A.}},
\bauthor{\bsnm{Kivshar}, \binits{Y.S.}}:
\batitle{Metasurfaces with {Maximum} {Chirality} {Empowered} by {Bound} {States} in the {Continuum}}.
\bjtitle{Physical Review Letters}
\bvolume{125}(\bissue{9}),
\bfpage{093903}
(\byear{2020})
\doiurl{10.1103/PhysRevLett.125.093903}
\end{barticle}
\endbibitem

\bibitem[\protect\citeauthoryear{Kühner et~al.}{2023}]{kuhner_unlocking_2023}
\begin{barticle}
\bauthor{\bsnm{Kühner}, \binits{L.}},
\bauthor{\bsnm{Wendisch}, \binits{F.J.}},
\bauthor{\bsnm{Antonov}, \binits{A.A.}},
\bauthor{\bsnm{Bürger}, \binits{J.}},
\bauthor{\bsnm{Hüttenhofer}, \binits{L.}},
\bauthor{\bsnm{De~S.~Menezes}, \binits{L.}},
\bauthor{\bsnm{Maier}, \binits{S.A.}},
\bauthor{\bsnm{Gorkunov}, \binits{M.V.}},
\bauthor{\bsnm{Kivshar}, \binits{Y.}},
\bauthor{\bsnm{Tittl}, \binits{A.}}:
\batitle{Unlocking the out-of-plane dimension for photonic bound states in the continuum to achieve maximum optical chirality}.
\bjtitle{Light: Science \& Applications}
\bvolume{12}(\bissue{1}),
\bfpage{250}
(\byear{2023})
\doiurl{10.1038/s41377-023-01295-z}
\end{barticle}
\endbibitem

\bibitem[\protect\citeauthoryear{Zhu et~al.}{2013}]{6529103}
\begin{barticle}
\bauthor{\bsnm{Zhu}, \binits{H.L.}},
\bauthor{\bsnm{Cheung}, \binits{S.W.}},
\bauthor{\bsnm{Chung}, \binits{K.L.}},
\bauthor{\bsnm{Yuk}, \binits{T.I.}}:
\batitle{Linear-to-circular polarization conversion using metasurface}.
\bjtitle{IEEE Transactions on Antennas and Propagation}
\bvolume{61}(\bissue{9}),
\bfpage{4615}--\blpage{4623}
(\byear{2013})
\doiurl{10.1109/TAP.2013.2267712}
\end{barticle}
\endbibitem

\bibitem[\protect\citeauthoryear{Shah et~al.}{2022}]{doi:10.1021/acsphotonics.2c00395}
\begin{barticle}
\bauthor{\bsnm{Shah}, \binits{Y.D.}},
\bauthor{\bsnm{Dada}, \binits{A.C.}},
\bauthor{\bsnm{Grant}, \binits{J.P.}},
\bauthor{\bsnm{Cumming}, \binits{D.R.S.}},
\bauthor{\bsnm{Altuzarra}, \binits{C.}},
\bauthor{\bsnm{Nowack}, \binits{T.S.}},
\bauthor{\bsnm{Lyons}, \binits{A.}},
\bauthor{\bsnm{Clerici}, \binits{M.}},
\bauthor{\bsnm{Faccio}, \binits{D.}}:
\batitle{An all-dielectric metasurface polarimeter}.
\bjtitle{ACS Photonics}
\bvolume{9}(\bissue{10}),
\bfpage{3245}--\blpage{3252}
(\byear{2022})
\doiurl{10.1021/acsphotonics.2c00395}
{\href{https://arxiv.org/abs/https://doi.org/10.1021/acsphotonics.2c00395}{{https://doi.org/10.1021/acsphotonics.2c00395}}}
\end{barticle}
\endbibitem

\bibitem[\protect\citeauthoryear{Han et~al.}{2024}]{han_chiral_2024}
\begin{botherref}
\oauthor{\bsnm{Han}, \binits{J.}},
\oauthor{\bsnm{Jang}, \binits{H.}},
\oauthor{\bsnm{Lim}, \binits{Y.}},
\oauthor{\bsnm{Kim}, \binits{S.}},
\oauthor{\bsnm{Lee}, \binits{J.}},
\oauthor{\bsnm{Jun}, \binits{Y.C.}}:
Chiral {Emission} from {Optical} {Metasurfaces} and {Metacavities}.
Advanced Photonics Research,
2400060
(2024)
\doiurl{10.1002/adpr.202400060}
\end{botherref}
\endbibitem

\bibitem[\protect\citeauthoryear{Sun et~al.}{2025}]{sun_circularly_2025}
\begin{barticle}
\bauthor{\bsnm{Sun}, \binits{K.}},
\bauthor{\bsnm{Yang}, \binits{B.}},
\bauthor{\bsnm{Cai}, \binits{Y.}},
\bauthor{\bsnm{Kivshar}, \binits{Y.}},
\bauthor{\bsnm{Han}, \binits{Z.}}:
\batitle{Circularly polarized thermal emission driven by chiral flatbands in monoclinic metasurfaces}.
\bjtitle{Science Advances}
\bvolume{11}(\bissue{31}),
\bfpage{0986}
(\byear{2025})
\doiurl{10.1126/sciadv.adw0986}
\end{barticle}
\endbibitem

\bibitem[\protect\citeauthoryear{Deng et~al.}{2025}]{deng_chiral_2025}
\begin{barticle}
\bauthor{\bsnm{Deng}, \binits{H.}},
\bauthor{\bsnm{Jiang}, \binits{X.}},
\bauthor{\bsnm{Zhang}, \binits{Y.}},
\bauthor{\bsnm{Zeng}, \binits{Y.}},
\bauthor{\bsnm{Barkaoui}, \binits{H.}},
\bauthor{\bsnm{Xiao}, \binits{S.}},
\bauthor{\bsnm{Yu}, \binits{S.}},
\bauthor{\bsnm{Kivshar}, \binits{Y.}},
\bauthor{\bsnm{Song}, \binits{Q.}}:
\batitle{Chiral lasing enabled by strong coupling}.
\bjtitle{Science Advances}
\bvolume{11}(\bissue{15}),
\bfpage{9562}
(\byear{2025})
\doiurl{10.1126/sciadv.ads9562}
\end{barticle}
\endbibitem

\bibitem[\protect\citeauthoryear{Heimig et~al.}{2026}]{heimig_chiral_2026}
\begin{barticle}
\bauthor{\bsnm{Heimig}, \binits{C.}},
\bauthor{\bsnm{Antonov}, \binits{A.A.}},
\bauthor{\bsnm{Gryb}, \binits{D.}},
\bauthor{\bsnm{Possmayer}, \binits{T.}},
\bauthor{\bsnm{Weber}, \binits{T.}},
\bauthor{\bsnm{Hirler}, \binits{M.}},
\bauthor{\bsnm{Biechteler}, \binits{J.}},
\bauthor{\bsnm{Sortino}, \binits{L.}},
\bauthor{\bsnm{De~S.~Menezes}, \binits{L.}},
\bauthor{\bsnm{Maier}, \binits{S.A.}},
\bauthor{\bsnm{Gorkunov}, \binits{M.V.}},
\bauthor{\bsnm{Kivshar}, \binits{Y.}},
\bauthor{\bsnm{Tittl}, \binits{A.}}:
\batitle{Chiral nonlinear polaritonics with van der {Waals} metasurfaces}.
\bjtitle{Science Advances}
\bvolume{12}(\bissue{13}),
\bfpage{5631}
(\byear{2026})
\doiurl{10.1126/sciadv.aeb5631}
\end{barticle}
\endbibitem

\bibitem[\protect\citeauthoryear{Yao et~al.}{2026}]{yao_kirigamibased_2026}
\begin{barticle}
\bauthor{\bsnm{Yao}, \binits{Y.}},
\bauthor{\bsnm{Kang}, \binits{S.}},
\bauthor{\bsnm{Luo}, \binits{A.}},
\bauthor{\bsnm{Yu}, \binits{J.}},
\bauthor{\bsnm{Qin}, \binits{K.}},
\bauthor{\bsnm{Zhang}, \binits{X.}},
\bauthor{\bsnm{Fan}, \binits{J.}},
\bauthor{\bsnm{Xia}, \binits{X.}},
\bauthor{\bsnm{Li}, \binits{H.}},
\bauthor{\bsnm{Wu}, \binits{X.}}:
\batitle{Kirigami‐{Based} {Flexible} {Metasurface} with {Reconfigurable} {Intrinsic} {Chirality} from {Zero} to {Near}‐{Unity}}.
\bjtitle{Advanced Optical Materials}
\bvolume{14}(\bissue{3}),
\bfpage{02199}
(\byear{2026})
\doiurl{10.1002/adom.202502199}
\end{barticle}
\endbibitem

\bibitem[\protect\citeauthoryear{Cen et~al.}{2025}]{cen_moire_2025}
\begin{barticle}
\bauthor{\bsnm{Cen}, \binits{M.}},
\bauthor{\bsnm{Wang}, \binits{J.}},
\bauthor{\bsnm{Cheng}, \binits{M.}},
\bauthor{\bsnm{Lei}, \binits{Z.}},
\bauthor{\bsnm{Li}, \binits{Y.}},
\bauthor{\bsnm{Wang}, \binits{Z.}},
\bauthor{\bsnm{Zhao}, \binits{X.}},
\bauthor{\bsnm{Wu}, \binits{Z.}},
\bauthor{\bsnm{Zhang}, \binits{H.}},
\bauthor{\bsnm{Liu}, \binits{Y.J.}}:
\batitle{Moiré metasurfaces with tunable near-infrared-{I} chiroptical responses for biomolecular chirality discrimination}.
\bjtitle{Nanoscale}
\bvolume{17}(\bissue{4}),
\bfpage{1970}--\blpage{1979}
(\byear{2025})
\doiurl{10.1039/D4NR03952A}
\end{barticle}
\endbibitem

\bibitem[\protect\citeauthoryear{Du et~al.}{2026}]{Du:26}
\begin{barticle}
\bauthor{\bsnm{Du}, \binits{F.}},
\bauthor{\bsnm{Tang}, \binits{H.}},
\bauthor{\bsnm{Liu}, \binits{Y.}},
\bauthor{\bsnm{Zhang}, \binits{M.}},
\bauthor{\bsnm{Lou}, \binits{B.}},
\bauthor{\bsnm{Gao}, \binits{G.}},
\bauthor{\bsnm{Li}, \binits{X.}},
\bauthor{\bsnm{Enriquez}, \binits{A.}},
\bauthor{\bsnm{Fan}, \binits{S.}},
\bauthor{\bsnm{Mazur}, \binits{E.}}:
\batitle{Dynamic control of intrinsic optical chirality via mems-integrated photonic crystals}.
\bjtitle{Optica}
\bvolume{13}(\bissue{3}),
\bfpage{449}--\blpage{456}
(\byear{2026})
\doiurl{10.1364/OPTICA.578880}
\end{barticle}
\endbibitem

\bibitem[\protect\citeauthoryear{Ji et~al.}{2021}]{ji_active_2021}
\begin{barticle}
\bauthor{\bsnm{Ji}, \binits{Y.}},
\bauthor{\bsnm{Fan}, \binits{F.}},
\bauthor{\bsnm{Zhang}, \binits{Z.}},
\bauthor{\bsnm{Tan}, \binits{Z.}},
\bauthor{\bsnm{Zhang}, \binits{X.}},
\bauthor{\bsnm{Yuan}, \binits{Y.}},
\bauthor{\bsnm{Cheng}, \binits{J.}},
\bauthor{\bsnm{Chang}, \binits{S.}}:
\batitle{Active terahertz spin state and optical chirality in liquid crystal chiral metasurface}.
\bjtitle{Physical Review Materials}
\bvolume{5}(\bissue{8}),
\bfpage{085201}
(\byear{2021})
\doiurl{10.1103/PhysRevMaterials.5.085201}
\end{barticle}
\endbibitem

\bibitem[\protect\citeauthoryear{Li et~al.}{2025}]{li_dynamic_2025}
\begin{barticle}
\bauthor{\bsnm{Li}, \binits{S.}},
\bauthor{\bsnm{Zhang}, \binits{Y.}},
\bauthor{\bsnm{Wang}, \binits{Y.}},
\bauthor{\bsnm{Cao}, \binits{G.}},
\bauthor{\bsnm{Liang}, \binits{Q.}},
\bauthor{\bsnm{Zhang}, \binits{X.}},
\bauthor{\bsnm{Sun}, \binits{H.}},
\bauthor{\bsnm{Zhang}, \binits{Y.}},
\bauthor{\bsnm{Wang}, \binits{Z.}},
\bauthor{\bsnm{Liu}, \binits{X.}},
\bauthor{\bsnm{Chen}, \binits{P.}},
\bauthor{\bsnm{Lin}, \binits{H.}},
\bauthor{\bsnm{Jia}, \binits{B.}},
\bauthor{\bsnm{Lu}, \binits{Y.-Q.}},
\bauthor{\bsnm{Li}, \binits{J.}}:
\batitle{Dynamic optical chirality based on liquid-crystal-embedded nano-cilia photonic structures}.
\bjtitle{Nature Communications}
\bvolume{16}(\bissue{1}),
\bfpage{6569}
(\byear{2025})
\doiurl{10.1038/s41467-025-61982-w}
\end{barticle}
\endbibitem

\bibitem[\protect\citeauthoryear{Sha et~al.}{2024}]{sha_chirality_2024}
\begin{barticle}
\bauthor{\bsnm{Sha}, \binits{X.}},
\bauthor{\bsnm{Du}, \binits{K.}},
\bauthor{\bsnm{Zeng}, \binits{Y.}},
\bauthor{\bsnm{Lai}, \binits{F.}},
\bauthor{\bsnm{Yin}, \binits{J.}},
\bauthor{\bsnm{Zhang}, \binits{H.}},
\bauthor{\bsnm{Song}, \binits{B.}},
\bauthor{\bsnm{Han}, \binits{J.}},
\bauthor{\bsnm{Xiao}, \binits{S.}},
\bauthor{\bsnm{Kivshar}, \binits{Y.}},
\bauthor{\bsnm{Song}, \binits{Q.}}:
\batitle{Chirality tuning and reversing with resonant phase-change metasurfaces}.
\bjtitle{Science Advances}
\bvolume{10}(\bissue{21}),
\bfpage{9017}
(\byear{2024})
\doiurl{10.1126/sciadv.adn9017}
\end{barticle}
\endbibitem

\bibitem[\protect\citeauthoryear{Koshelev et~al.}{2020}]{koshelev_engineering_2020}
\begin{barticle}
\bauthor{\bsnm{Koshelev}, \binits{K.}},
\bauthor{\bsnm{Bogdanov}, \binits{A.}},
\bauthor{\bsnm{Kivshar}, \binits{Y.}}:
\batitle{Engineering with {Bound} {States} in the {Continuum}}.
\bjtitle{Optics and Photonics News}
\bvolume{31}(\bissue{1}),
\bfpage{38}
(\byear{2020})
\doiurl{10.1364/OPN.31.1.000038}
\end{barticle}
\endbibitem

\bibitem[\protect\citeauthoryear{Aigner et~al.}{2025}]{aigner2025optical}
\begin{barticle}
\bauthor{\bsnm{Aigner}, \binits{A.}},
\bauthor{\bsnm{Possmayer}, \binits{T.}},
\bauthor{\bsnm{Weber}, \binits{T.}},
\bauthor{\bsnm{Antonov}, \binits{A.A.}},
\bauthor{\bsnm{S.~Menezes}, \binits{L.}},
\bauthor{\bsnm{Maier}, \binits{S.A.}},
\bauthor{\bsnm{Tittl}, \binits{A.}}:
\batitle{Optical control of resonances in temporally symmetry-broken metasurfaces}.
\bjtitle{Nature}
\bvolume{644}(\bissue{8078}),
\bfpage{896}--\blpage{902}
(\byear{2025})
\doiurl{10.1038/s41586-025-09363-7}
\end{barticle}
\endbibitem

\bibitem[\protect\citeauthoryear{Brikh et~al.}{2025}]{brikh2025mid}
\begin{barticle}
\bauthor{\bsnm{Brikh}, \binits{F.U.}},
\bauthor{\bsnm{Ezerskii}, \binits{A.}},
\bauthor{\bsnm{Pashina}, \binits{O.}},
\bauthor{\bsnm{Glebov}, \binits{N.}},
\bauthor{\bsnm{Yin}, \binits{W.}},
\bauthor{\bsnm{Makarov}, \binits{S.V.}},
\bauthor{\bsnm{Petrov}, \binits{M.}},
\bauthor{\bsnm{Sinev}, \binits{I.}},
\bauthor{\bsnm{Altug}, \binits{H.}}:
\batitle{Mid-ir light modulators enabled by dynamically tunable ultra high-q silicon membrane metasurfaces}.
\bjtitle{arXiv preprint arXiv:2509.23167}
(\byear{2025})
\doiurl{10.48550/arXiv.2509.23167}
\end{barticle}
\endbibitem

\bibitem[\protect\citeauthoryear{Gorkunov et~al.}{2025}]{GorkunovMaxim_AOM}
\begin{barticle}
\bauthor{\bsnm{Gorkunov}, \binits{M.V.}},
\bauthor{\bsnm{Antonov}, \binits{A.A.}},
\bauthor{\bsnm{Mamonova}, \binits{A.V.}},
\bauthor{\bsnm{Muljarov}, \binits{E.A.}},
\bauthor{\bsnm{Kivshar}, \binits{Y.}}:
\batitle{Substrate-induced maximum optical chirality of planar dielectric structures}.
\bjtitle{Advanced Optical Materials}
\bvolume{13}(\bissue{3}),
\bfpage{2402133}
(\byear{2025})
\doiurl{10.1002/adom.202402133}
\end{barticle}
\endbibitem

\bibitem[\protect\citeauthoryear{Sakoda}{1995}]{sakoda_symmetry_1995}
\begin{barticle}
\bauthor{\bsnm{Sakoda}, \binits{K.}}:
\batitle{Symmetry, degeneracy, and uncoupled modes in two-dimensional photonic lattices}.
\bjtitle{Physical Review B}
\bvolume{52}(\bissue{11}),
\bfpage{7982}--\blpage{7986}
(\byear{1995})
\doiurl{10.1103/PhysRevB.52.7982} .
Accessed 2023-09-12
\end{barticle}
\endbibitem

\bibitem[\protect\citeauthoryear{Hopkins et~al.}{2016}]{hopkins_circular_2016}
\begin{barticle}
\bauthor{\bsnm{Hopkins}, \binits{B.}},
\bauthor{\bsnm{Poddubny}, \binits{A.N.}},
\bauthor{\bsnm{Miroshnichenko}, \binits{A.E.}},
\bauthor{\bsnm{Kivshar}, \binits{Y.S.}}:
\batitle{Circular dichroism induced by {Fano} resonances in planar chiral oligomers}.
\bjtitle{Laser \& Photonics Reviews}
\bvolume{10}(\bissue{1}),
\bfpage{137}--\blpage{146}
(\byear{2016})
\doiurl{10.1002/lpor.201500222} .
\bcomment{arXiv: 1412.1120}.
Accessed 2021-10-30
\end{barticle}
\endbibitem

\bibitem[\protect\citeauthoryear{Kondratov et~al.}{2016}]{kondratov_extreme_2016}
\begin{barticle}
\bauthor{\bsnm{Kondratov}, \binits{A.V.}},
\bauthor{\bsnm{Gorkunov}, \binits{M.V.}},
\bauthor{\bsnm{Darinskii}, \binits{A.N.}},
\bauthor{\bsnm{Gainutdinov}, \binits{R.V.}},
\bauthor{\bsnm{Rogov}, \binits{O.Y.}},
\bauthor{\bsnm{Ezhov}, \binits{A.A.}},
\bauthor{\bsnm{Artemov}, \binits{V.V.}}:
\batitle{Extreme optical chirality of plasmonic nanohole arrays due to chiral {Fano} resonance}.
\bjtitle{Physical Review B}
\bvolume{93}(\bissue{19}),
\bfpage{195418}
(\byear{2016})
\doiurl{10.1103/physrevb.93.195418}
\end{barticle}
\endbibitem

\bibitem[\protect\citeauthoryear{Overvig et~al.}{2018}]{overvig_dimerized_2018}
\begin{barticle}
\bauthor{\bsnm{Overvig}, \binits{A.C.}},
\bauthor{\bsnm{Shrestha}, \binits{S.}},
\bauthor{\bsnm{Yu}, \binits{N.}}:
\batitle{Dimerized high contrast gratings}.
\bjtitle{Nanophotonics}
\bvolume{7}(\bissue{6}),
\bfpage{1157}--\blpage{1168}
(\byear{2018})
\doiurl{10.1515/nanoph-2017-0127} .
Accessed 2022-11-05
\end{barticle}
\endbibitem

\bibitem[\protect\citeauthoryear{Wang et~al.}{2023}]{wang_brillouin_2023}
\begin{barticle}
\bauthor{\bsnm{Wang}, \binits{W.}},
\bauthor{\bsnm{Srivastava}, \binits{Y.K.}},
\bauthor{\bsnm{Tan}, \binits{T.C.}},
\bauthor{\bsnm{Wang}, \binits{Z.}},
\bauthor{\bsnm{Singh}, \binits{R.}}:
\batitle{Brillouin zone folding driven bound states in the continuum}.
\bjtitle{Nature Communications}
\bvolume{14}(\bissue{1}),
\bfpage{2811}
(\byear{2023})
\doiurl{10.1038/s41467-023-38367-y} .
Accessed 2025-06-18
\end{barticle}
\endbibitem

\bibitem[\protect\citeauthoryear{Sivan and Spector}{2020}]{sivan2020ultrafast}
\begin{barticle}
\bauthor{\bsnm{Sivan}, \binits{Y.}},
\bauthor{\bsnm{Spector}, \binits{M.}}:
\batitle{Ultrafast dynamics of optically induced heat gratings in metals}.
\bjtitle{ACS Photonics}
\bvolume{7}(\bissue{5}),
\bfpage{1271}--\blpage{1279}
(\byear{2020})
\doiurl{10.1021/acsphotonics.0c00224}
\end{barticle}
\endbibitem

\bibitem[\protect\citeauthoryear{Toftul et~al.}{2024}]{toftul_chiral_2024}
\begin{barticle}
\bauthor{\bsnm{Toftul}, \binits{I.}},
\bauthor{\bsnm{Tonkaev}, \binits{P.}},
\bauthor{\bsnm{Koshelev}, \binits{K.}},
\bauthor{\bsnm{Lai}, \binits{F.}},
\bauthor{\bsnm{Song}, \binits{Q.}},
\bauthor{\bsnm{Gorkunov}, \binits{M.}},
\bauthor{\bsnm{Kivshar}, \binits{Y.}}:
\batitle{Chiral {Dichroism} in {Resonant} {Metasurfaces} with {Monoclinic} {Lattices}}.
\bjtitle{Physical Review Letters}
\bvolume{133}(\bissue{21}),
\bfpage{216901}
(\byear{2024})
\doiurl{10.1103/PhysRevLett.133.216901}
\end{barticle}
\endbibitem

\bibitem[\protect\citeauthoryear{Fedotov et~al.}{2006}]{fedotov2006asymmetric}
\begin{barticle}
\bauthor{\bsnm{Fedotov}, \binits{V.}},
\bauthor{\bsnm{Mladyonov}, \binits{P.}},
\bauthor{\bsnm{Prosvirnin}, \binits{S.}},
\bauthor{\bsnm{Rogacheva}, \binits{A.}},
\bauthor{\bsnm{Chen}, \binits{Y.}},
\bauthor{\bsnm{Zheludev}, \binits{N.}}:
\batitle{Asymmetric propagation of electromagnetic waves through a planar chiral structure}.
\bjtitle{Physical review letters}
\bvolume{97}(\bissue{16}),
\bfpage{167401}
(\byear{2006})
\doiurl{10.1103/PhysRevLett.97.167401}
\end{barticle}
\endbibitem

\bibitem[\protect\citeauthoryear{Sokolowski-Tinten and Von~der Linde}{2000}]{sokolowski2000generation}
\begin{barticle}
\bauthor{\bsnm{Sokolowski-Tinten}, \binits{K.}},
\bauthor{\bsnm{Linde}, \binits{D.}}:
\batitle{Generation of dense electron-hole plasmas in silicon}.
\bjtitle{Physical Review B}
\bvolume{61}(\bissue{4}),
\bfpage{2643}
(\byear{2000})
\doiurl{10.1103/PhysRevB.61.2643}
\end{barticle}
\endbibitem

\bibitem[\protect\citeauthoryear{Karl et~al.}{2020}]{KarlNanoLett}
\begin{barticle}
\bauthor{\bsnm{Karl}, \binits{N.}},
\bauthor{\bsnm{Vabishchevich}, \binits{P.P.}},
\bauthor{\bsnm{Shcherbakov}, \binits{M.R.}},
\bauthor{\bsnm{Liu}, \binits{S.}},
\bauthor{\bsnm{Sinclair}, \binits{M.B.}},
\bauthor{\bsnm{Shvets}, \binits{G.}},
\bauthor{\bsnm{Brener}, \binits{I.}}:
\batitle{Frequency conversion in a time-variant dielectric metasurface}.
\bjtitle{Nano Letters}
\bvolume{20}(\bissue{10}),
\bfpage{7052}--\blpage{7058}
(\byear{2020})
\doiurl{10.1021/acs.nanolett.0c02113} .
\bcomment{PMID: 32940476}
\end{barticle}
\endbibitem

\end{thebibliography}

\section*{Acknowledgements}
We acknowledge the European Union's Horizon Europe Research and Innovation Programme under agreements 101046424 (TwistedNano) and 101070700 (MIRAQLS). This work was supported by the Swiss State Secretariat for Education, Research and Innovation (SERI) under contract numbers 22.00018 and 22.00081. The authors acknowledge the use of nanofabrication facilities at the Center of MicroNano Technology of \'Ecole Polytechnique F\'ed\'erale de Lausanne. O.P. and M.P acknowledge the the support of the work by the Federal Academic Leadership Program Priority 2030. This research was also supported financially by the Overseas Outstanding Youth of Shandong Province (Grants 2024HWYQ-082). S.M. acknowledges the support of National Natural Science Foundation of China (project 62350610272) and the Department of Science and Technology of Shandong Province (Grant KY0020240040). The work of A.M. was carried out within
the State assignment of NRC ``Kurchatov Institute''.
M.G. acknowledges the support from the Federal Academic Leadership Program Priority 2030 (NUST MISIS Strategic Technology Project ``Quantum Internet'')

\section*{Competing Interests Statement}
The authors declare no competing interests.


\end{document}